\documentclass[a4paper,12pt]{article}
\pdfoutput=1
\usepackage{epsfig}
\usepackage{amsmath}
\usepackage{amssymb}
\usepackage{amsfonts}
\usepackage{bbm}
\usepackage{fleqn}
\usepackage[footnotesize]{caption}
\usepackage{graphicx}
\usepackage{mathrsfs}
\usepackage[center,footnotesize,hang]{subfigure}

\def\be{\begin{equation}}
\def\ee{\end{equation}}
\def\beq{\begin{equation}}
\def\eeq{\end{equation}}
\def\bea{\begin{eqnarray}}
\def\eea{\end{eqnarray}}

\def\<{\left\langle}
\def\>{\right\rangle}

\allowdisplaybreaks[2]
\begin{document}
\bibliographystyle{OurBibTeX}
\begin{titlepage}
 \vspace*{-15mm}
\begin{flushright}
%hep-ph/yymmnnn\\
\end{flushright}
\vspace*{5mm}
\begin{center}
{ \sffamily \LARGE Implications of large CP Violation in B mixing for Supersymmetric Standard Models}
\\[8mm]
S.~F.~King\footnote{E-mail:
\texttt{king@soton.ac.uk}}
\\[3mm]
{\small\it
School of Physics and Astronomy,
University of Southampton,\\
Southampton, SO17 1BJ, U.K.
}\\[1mm]
\end{center}
\vspace*{0.75cm}
\begin{abstract}
\noindent
Following the anomalous like-sign dimuon charge asymmetry measured by the D0 collaboration 
at the Tevatron collider we discuss the implications of large CP violation in $B_{d,s}$ mixing for Supersymmetric (SUSY) Standard Models, focussing on those models which allow a family symmetry and unification. For the Minimal Supersymmetric Standard Model (MSSM) we show that it is only possible to account for $B_{s}$ 
mixing and CP violation at the expense of large squark mixing which would require a 
new approach to family symmetry models. 
In order to describe both $B_{s}$ and $B_{d}$ mixing and CP violation we are led to consider 
SUSY models with Higgs fields transforming as triplets under a family symmetry. We describe a realistic
such model based on $\Delta_{27}$ family symmetry
in which tree-level exchange of the second Higgs family predicts $B_{s}$ and $B_{d}$ mixing and CP violation in good agreement with a recent global fit, while naturally suppressing flavour and CP violation involving the first and second quark and lepton families.
\end{abstract}
\end{titlepage}
\newpage
\setcounter{footnote}{0}

\section{Introduction}

The Standard Model (SM) has provided a remarkably successful description of quarks and leptons 
from its inception in the 1960s until the end of the last millennium. In 1998 the discovery of neutrino mass and mixing demanded new physics beyond the SM for its explanation
(for a review see e.g. \cite{Mohapatra:2005wg}). The picture which has emerged in the lepton sector is consistent with three neutrino mass and mixing described by a PMNS matrix $U$, although its origin remains unclear and $U_{e3}$ is so far unmeasured. By contrast, despite intense experimental and theoretical scrutiny, there has been no firm evidence of any new physics in the quark sector, with the CKM picture of CP violation, summarised by the Unitarity Triangle, becoming ever more precisely determined \cite{Amsler:2008zzb}. Yet despite this progress, some cracks have begun to appear in B physics which may call for new physics to describe CP violation beyond the CKM matrix. It is worth recalling that CP violation is predicted by the SM to be very small in $B-\overline{B}$ mixing, well below the Tevatron sensitivity,
due in part to the small phases of the relevant CKM elements. On the other hand new physics can compete with the SM box diagrams, in principle with large new CP violating phases, giving much larger CP asymmetries in $B-\overline{B}$ mixing than predicted by the SM, rendering it observable at the Tevatron.

Recently the D0 Collaboration has reported evidence 
for CP violation in the like-sign
dimuon charge asymmetry \cite{Abazov:2010hv} 
\begin{eqnarray}
A_{\rm sl}^b \equiv \frac{N_b^{++}-N_b^{--}}{N_b^{++}+N_b^{--}}
 = - (0.957 \pm 0.251 \pm 0.146)\times 10^{-2},
\end{eqnarray}
where $N_b^{++}\;(N^{--}_b)$ is the number of events with $b\; (\bar b)$ containing hadrons decaying semileptonically into $\mu^+ X\;(\mu^- X)$ .  The D0 result
is 3.2$\sigma$ away from the standard model (SM) prediction 
$(-2.3 \pm 0.5)\times 10^{-4}$   \cite{Lenz:2006hd}.
The CDF   \cite{cdf} measurement of $A_{\rm sl}^b$, using only 1.6 fb$^{-1}$ of data, 
as compared to the D0 6.1 fb$^{-1}$ data set,
has a central value which is positive,
$A_{\rm sl}^b = (8.0 \pm 9.0 \pm 6.8) \times 10^{-3} $, but is still compatible with the D0 measurement at the 1.5$\sigma$ level
because its uncertainties are 4 times larger than those of D0. Combining the
the D0 and CDF results for $A_{\rm sl}^b$,  one finds $A_{\rm sl}^b \simeq - (0.85 \pm 0.28)\times 10^{-2}$ which is still 3$\sigma$ away from the SM value. 

The interpretation of the observed CP asymmetry is in terms of the production
of  $B\overline{B}$ meson pairs followed by their subsequent oscillation and 
semi-leptonic decay where the charge of the final state lepton effectively tags 
whether it is a $b$ or $\overline{b}$ quark which decays.
Thus the dilepton asymmetry can be written as 
\begin{eqnarray}
A_{\rm sl}^b \equiv \frac{N(BB)-N(\overline{B} \overline{B})}{N(BB)+N(\overline{B} \overline{B})}
=  \frac{
\frac{P_{\overline{B}\rightarrow B}}
{P_{\overline{B}\rightarrow \overline{B}}}
-\frac{P_{B\rightarrow \overline{B}}}{P_{B\rightarrow B}} }
{\frac{P_{\overline{B}\rightarrow B}}{P_{\overline{B}\rightarrow \overline{B}}}
+\frac{P_{B\rightarrow \overline{B}}}{P_{B\rightarrow B}} }.
\end{eqnarray}
The measured asymmetry at the Tevatron is interpreted as a linear combination of  the asymmetries 
$a^{d,s}_{\rm sl}$ in $B_d$ and $B_s$ oscillation and decays \cite{Abazov:2010hv},
\begin{eqnarray}
A^b_{\rm sl}&=& (0.506\pm 0.043) a^d_{\rm sl} + (0.494\pm 0.043)a^s_{\rm sl},
\end{eqnarray}
where the ``wrong charge'' asymmetries are 
\be
a_{\rm sl}^q \equiv \frac{\Gamma(\bar{B}_q \to \mu^+ X)-\Gamma(B_q \to \mu^-X)}{\Gamma(\bar{B}_q \to \mu^+ X)+\Gamma(B_q \to \mu^-X)} ~~.
\ee
The current experimental values of the separate asymmetries are  
$a_{\rm sl}^d = -(0.47\pm 0.46) \times 10^{-2} $ \cite{Barberio:2008fa} and  
$a_{\rm sl}^s = - (0.17 \pm 0.91 \pm 0.15) \times 10^{-2}$  \cite{Abazov:2009wg} which, while being consistent with a negative $A^b_{\rm sl}$ of order one per cent, are also consistent with zero as well, and so do not shed much light on which of the two separate asymmetries is responsible, however there is apparently a mild tendency for both of these asymmetries to be acting together.

In addition to these measurements, D0 have reconstructed $B_s\rightarrow J/\psi \phi $
decays and have measured the time dependent asymmetry parameter 
$S_{\psi \phi}=-\sin \phi_{B_s}$, where $\phi_{B_s}$ is the phase of the 
$B_s-\overline{B}_s$ mixing matrix element $M^{12}_s=|M^{12}_s|e^{i\phi_{B_s}}$, 
and finds a discrepancy with the SM prediction of $S_{\psi \phi}\sim 0$ 
at the level of $2.1\sigma$  \cite{:2008fj}.
A recent preliminary CDF analysis based on 5.2 fb$^{-1}$ of data, 
finds $S_{\psi \phi}$ which is consistent with zero but has a central value of 
$S_{\psi \phi}\sim 0.5$ \cite{CDF} .
The predictions for $S_{\psi \phi}$ in various models and the correlations between this and other observables 
as a means of discriminating between these models has been comprehensively
studied \cite{Altmannshofer:2009ne}.
Neglecting the small SM contribution to $S_{\psi \phi}$, the following model independent relation holds
between  $S_{\psi \phi}$ and $a_{\rm sl}^s$
\cite{Grossman:2009mn}:
\be
a_{\rm sl}^s \approx -\frac{|\Gamma_{12}^s|}{|M^{12}_{s}|}S_{\psi \phi}
%\frac{S_{\psi \phi}}{\sqrt{1-S_{\psi \phi}^2}}.
\label{a0}
\ee
Following the recent D0 results, it has been shown that it is possible to fit $S_{\psi \phi}$  
and $a_{\rm sl}^s$ from the latest data  by 
assuming that new physics contributes significantly to the mixing matrix element $M_{12}^s$ and also by
allowing the decay matrix element $\Gamma_{12}^s$ to float \cite{Ligeti:2010ia}.
More precisely the authors in \cite{Ligeti:2010ia} perform a global fit of all experimental measurements, including the recent D0 asymmetry results and the recent 
CDF preliminary results for $B_s\rightarrow J/\psi \phi $, allowing the two
SM decay matrix elements $\Gamma_{12}^q$ to float, while allowing for new physics to contribute to both mixing matrix elements $M_{12}^q$ which can be parameterised as:
\be
M^{12}_q= M^{12,SM}_{q}(1+h_qe^{i \theta_q}).
\label{h}
\ee
Using the convention where $a_{\rm sl}^q = \mathrm{Im}(-\Gamma_{12}^q/M^{12}_q)$ with the dominant real parts of $\Gamma_{12}^s$ and $M_{12}^s$ being positive in the SM, gives,
\be
a_{\rm sl}^s \approx \frac{|\Gamma_{12}^s|}{|M^{12,SM}_{s}|}
\frac{h_s\sin \theta_s}{1+2h_s\cos \theta_s + h_s^2},
\label{a}
\ee
 where we have neglected the small phase in the SM matrix element, $\beta_s\approx -0.01$.
However the corresponding SM matrix element in the $B_d$ sector is,
$M^{12,SM}_{d}=|M^{12,SM}_{d}|e^{2i \beta}$, with $\beta \approx 0.38$, which will contribute
significantly to the phase of $M^{12}_d$.

The global fit \cite{Ligeti:2010ia} 
includes the measured $B_s$ and $B_d$ mass differences $\Delta M_s$ and 
$\Delta M_d$, the measured time dependent asymmetries $S_{\psi \phi}$,
$S_{\psi K}$ (which determines the unitarity triangle angle $\beta$), the
CP asymmetries $a^{d,s}_{\rm sl}$ as well as the CKM parameters 
$\overline{\rho}, \overline{\eta}$, while allowing $\Gamma_{12}^q$ to vary from its
SM predicted value.
The best fit value for $\Gamma_{12}^s$ is about twice as large as the SM prediction \cite{Lenz:2006hd},
with similar results obtained in \cite{Bauer:2010dg} where the implications of such a large value 
are discussed. The decay matrix element $\Gamma_{12}^s$ is proportional to the square of a tree-level SM amplitude proportional to $V_{cb}$ arising from $W$ exchange so it is challenging to understand why the best fit value should be so large.

The best fit points for new physics contributions to the matrix elements are \cite{Ligeti:2010ia}
\be
(h_d, \theta_d)\sim (0.25, \frac{9}{8}\pi), \ \ (h_s, \theta_s)_{I}\sim (0.6, \frac{11}{8}\pi), \ \ 
(h_s, \theta_s)_{II}\sim (1.9, \frac{9}{8}\pi). 
\label{bestfit}
\ee
Note that there are two different best fit points for $(h_s, \theta_s)$, but only a single best fit point for  
$(h_d, \theta_d)$ which is only preferred from a zero value at a confidence level of order 1.5$\sigma$.
All points have the angle $\theta_q$ 
in the third quadrant where both  $\sin \theta_q$ and $\cos \theta_q$ are negative, resulting
in enhanced negative values of $a_{\rm sl}^q$ from Eq.\ref{a}.
Although the precise best fit points must be regarded as indicative values with rather large error bars, 
the fit is quite robust with $h_s=h_d=0$ disfavoured at 3.3$\sigma$ \cite{Ligeti:2010ia}.
There have already been several attempts to explain the recent data 
 \cite{Dighe:2010nj,Dobrescu:2010rh,Chen:2010wv,Buras:2010mh,Chen:2010aq,Parry:2010ce,Ko:2010mn}
(see also \cite{Kawashima:2009jv}). 
 
In this paper we consider the implications of large CP violation in B mixing for SUSY Standard Models  (for a review see e.g. \cite{Chung:2003fi}) focussing on those models which include a family symmetry and allow for unification. 
We discuss two distinct possibilities. For the Minimal Supersymmetric Standard Model (MSSM), discussed in Section~2, we show that it is only possible to account for $B_{s}$ 
mixing and CP violation at the expense of large squark mixing which would require a 
new approach to family symmetry models. Moreover this approach cannot account for $B_d$ mixing and CP violation. In order to describe both $B_{s}$ and $B_{d}$ mixing and CP violation, in Section~3
we are led to consider SUSY models with Higgs fields transforming as triplets under a family symmetry, where tree-level exchange of the second Higgs family may readily account for all the data. We describe a realistic
such model based on $\Delta_{27}$ family symmetry \cite{Howl:2009ds}
which predicts $B_{s}$ and $B_{d}$ mixing and CP violation in good agreement with the best fit point I, and naturally leads to small effects in $K^0$ mixing and other flavour violating processes.

\section{Large CP Violation for $B_s$ mixing in the MSSM}

As discussed in \cite{Randall:1998te} the MSSM contributions to CP violation in B mixing arise dominantly from box diagrams involving down squarks and gluinos. The $B_q$ mixing matrix element can be written as the sum of the SM box diagrams and the SUSY box diagrams,
\be
M^{12}_q = M^{12,\mathrm{SM}}_{q} + M^{12,\mathrm{SUSY}}_{q}
\ee
where from Eq.\ref{h} we identify
\be
h_qe^{i \theta_q}=\frac{ M^{12,\mathrm{SUSY}}_{q}}{M^{12,\mathrm{SM}}_{q}},
\ee
where $q=s,d$.
Using the matrix elements in \cite{Randall:1998te} with the updated parameters in \cite{Buras:2010mh} we find, in the mass insertion approximation (see \cite{Ciuchini:2007ha} and references therein),
\bea
h_se^{i \theta_s} & \approx &  \left( \frac{500\ \mathrm{GeV}}{m_{\tilde{q}} }\right) ^2
\left[  5 \left(  (\delta^d_{32})_{LL}^2 +  (\delta^d_{32})_{RR}^2 \right)   
- 190 \left( (\delta^d_{32})_{LL} (\delta^d_{32})_{RR}\right)  \right]   \label{SUSY}\\
h_de^{i ( \theta_d + 2\beta )} & \approx &  \left( \frac{500\ \mathrm{GeV}}{m_{\tilde{q}} }\right) ^2
\left[  114 \left(  (\delta^d_{31})_{LL}^2 +  (\delta^d_{31})_{RR}^2 \right)   
- 4460 \left( (\delta^d_{31})_{LL} (\delta^d_{31})_{RR}\right)  \right] , \nonumber
\eea
where $ (\delta^d_{3i})_{LL}=(V_{D_L}m^2_{\tilde{Q}}V_{D_L}^{\dagger})_{3i}/m_{\tilde{q}}^2$ and 
$ (\delta^d_{3i})_{RR}=(V_{D_R}m^2_{\tilde{D}}V_{D_R}^{\dagger})_{3i}/m_{\tilde{q}}^2$,
with $m_{\tilde{q}}$ being a typical squark mass (assumed to be degenerate with the gluino
$\tilde{g}$)
where $m^2_{\tilde{Q}}$ is the left-handed (L) squark doublet mass squared matrix, 
$m^2_{\tilde{D}}$ is the right-handed (R) down-type squark mass squared matrix, 
and $V_{D_L}$ and $V_{D_R}$ are the unitary matrices that diagonalise the 
down-type quark mass matrix $M^d$, 
namely $V_{D_L}M^dV_{D_R}^{\dagger} = \mathrm{diag}(m_d, m_s, m_b)$.
This is summarised by the statement that 
the LL and RR mass mixing between down squarks of different generations in the super CKM basis is the source of the flavour and CP violation \cite{Chung:2003fi}. We have not included the 
contributions from LR mass mixing which are tightly constrained by $b\rightarrow s \gamma$.

Comparing the SUSY predictions in Eq.\ref{SUSY} to the best fit points in Eq.\ref{bestfit}, it is clear that the values of $h_s \sim 1$ could either be achieved, assuming squark and gluino masses of 
about 500 GeV,  by $(\delta^d_{32})_{LL}\sim (\delta^d_{32})_{RR} \sim 0.05 -  0.1$
or $(\delta^d_{32})_{LL}\ll (\delta^d_{32})_{RR} \sim 0.3 -  0.6$ or 
 $(\delta^d_{32})_{RR}\ll (\delta^d_{32})_{LL} \sim 0.3 -  0.6$. These represent quite sizeable squark mixing angles,
which run into conflict with grand unified theories (GUTs) based on gravity mediated SUSY breaking. To see this, it is worth bearing in mind that renormalisation group (RG) running from the high energy GUT or Planck scale to low energies tends to increase the diagonal squark masses $m_{\tilde{q}}$ by about a factor of 5, while not enhancing the off-diagonal squark masses \cite{Ciuchini:2007ha}, so the high energy $ (\delta^d_{32})_{LL,RR}$ parameters need to be 25 times larger  than these low energy values which is not possible (they can at most only be of order unity). For some gauge mediated SUSY breaking scenario, where the messenger scale is below the GUT scale, the effect of running is reduced so it may be possible to achieve these low energy values. Another constraint is that, in the framework of GUTs, there is the danger of running into conflict with the bound on $ (\delta^e_{32})_{LL}<0.12$ from $\tau \rightarrow \mu \gamma$ \cite{Ciuchini:2007ha} since the slepton masses are only enhanced by about a factor of 2 in running from the GUT scale to low energies. In the context of $SU(5)$ GUTs 
the low energy $ (\delta^d_{32})_{LL}$ parameters are constrained by the high energy requirement that 
$ (\delta^d_{32})^{\mathrm{GUT}}_{RR}=  (\delta^e_{32})^{\mathrm{GUT}}_{LL}<0.48$  \cite{Ciuchini:2007ha},
which implies the low energy constraint  $(\delta^d_{32})_{RR}<0.02 $.  From the point of view of family $SU(3)$
symmetry models (see \cite{Antusch:2008jf} and references therein) 
small squark mixing parameters are also expected
$(\delta^d_{32})_{RR}\sim 10^{-3}$. It seems that the data is not favoured by conventional SUSY GUTs and
family symmetry. 

Suppose we abandon all pre-conceived prejudices about SUSY breaking, 
GUTs and family symmetry, but continue to assume that the MSSM is the only source of new physics. Then we can ask if it is possible for the MSSM to describe the observations and if so then what the data is telling us about squark mixing. For the reasons outlined above, from the point of view of the MSSM, it is desirable for the squark mixing to be as small as possible. Therefore we shall consider the smallest mixing describing the data given by solution I with 
$(\delta^d_{32})_{LL}\sim (\delta^d_{32})_{RR} \sim 0.05$. Taking into account the RG running up to the (unknown) SUSY breaking
messenger scale, this still suggests a high energy theory capable of giving quite large (2,3) mixing in the squark sector. Following this reasoning we are led to consider high energy Yukawa matrices in the quark sector which have a democratic structure in the (2,3) sectors,
\be
Y^u_{(2,3)} \sim Y^d_{(2,3)} \sim \left(
  \begin{matrix} % or pmatrix or bmatrix or Bmatrix or ...
  1 & 1 \\
      1 & 1 \\
   \end{matrix}
   \right).
\ee
This ensures that even approximately diagonal squark mass squared matrices would generate
large squark mixing in the super CKM basis. However the small CKM angle $\theta_{23} \sim |\theta^u_{23}- \theta^d_{23}|$
would then require an accurate cancellation. To enforce this (approximate) cancellation in a natural way we shall require
$Y^u_{(2,3)} \propto Y^d_{(2,3)}$ and rank one sub-matrices to achieve the (2,3) quark mass hierarchies.
 Both these features could be achieved by an $SU(3)$ 
family symmetry under which the left and right-handed quarks transform as triplets
$Q_i, U^c_i, D^c_i \sim 3$ and which is spontaneously
 broken by an anti-triplet 
 flavon $\phi_{23}^i$ with aligned vacuum expectation values 
 $\langle \phi_{23} \rangle \sim (0, 1, 1)V$, where phases    
 have been suppressed (see \cite{Antusch:2008jf}). Then the (2,3) 
 block could be generated from leading order operators of the form, dropping 
 coefficients,
 \be
 H^uQ_i\phi_{23}^i   U^c_j\phi_{23}^j   + H^dQ_i\phi_{23}^i   D^c_j\phi_{23}^j ,
 \label{o1}
  \ee 
which, after the two Higgs doublets of the MSSM $H^u$ and $H^d$ acquire their VEVs,
implies $\theta_{23}\sim 0$ and $m_s\ll m_b$,  $m_c\ll m_t$, at leading order.
This differs from the usual $SU(3)$ models  \cite{Antusch:2008jf} by the absence of the flavon 
$\phi_{3}^i$ with VEV $\langle \phi_{3} \rangle \sim (0, 0, 1)V_{23}$.
This democratic (2,3) structure could be plausibly be extended to the
charged lepton sector as well,
\be
Y^e_{(2,3)} \sim \left(
  \begin{matrix} % or pmatrix or bmatrix or Bmatrix or ...
  1 & 1 \\
      1 & 1 \\
   \end{matrix}
   \right),
\ee
resulting from the leading order operator,
\be
 H^dL_i\phi_{23}^i   E^c_j\phi_{23}^j ,
\label{o2}
  \ee
  which implies maximal charged lepton mixing in the (2,3) sector.
  In order to achieve maximal (2,3) physical lepton mixing we require the light effective neutrino Majorana mass matrix to be approximately diagonal.
For example this could be due to a type II see-saw mechanism via a sextet flavon $\Delta^{ij}$ with 
  an approximately diagonal VEV $\langle \Delta^{ij} \rangle \sim \mathrm{diag}(0,a,b)$ arising from 
  operators of the form \cite{King:2009tj},
  \be
  H^uH^uL_i L_j \Delta^{ij}.
   \label{03}
      \ee
 This then leads to maximal atmospheric neutrino mixing coming from the charged lepton sector.
 
  In the above approach we are essentially saturating the high energy limits with 
  $ (\delta^d_{32})^{\mathrm{GUT}}_{LL,RR}\sim   (\delta^e_{32})^{\mathrm{GUT}}_{LL,RR} \sim 1$ leading to some apparent tension with the  previously mentioned 
  $\tau \rightarrow \mu \gamma$ limit  $ (\delta^e_{32})^{\mathrm{GUT}}_{LL} <0.48 $.
  However there is a very large compensating effect from RG running due to the large charged lepton Yukawa couplings which will tend to reduce the magnitude of the slepton masses, including the off-diagonal ones.
  To leading log, the correction will be 
  \be
  \Delta (\delta^e_{32})_{LL} \sim -\frac{1}{8\pi^2}\ln \frac{M_{GUT}}{M_W}\sim -0.4
  \ee
  effectively relaxing the tension and allowing $ (\delta^e_{32})^{\mathrm{GUT}}_{LL} \sim 1 $.  
  Nevertheless,  $\tau \rightarrow \mu \gamma$ might be expected to be not far below its current limit. Such Yukawa induced RG corrections will also be present for the squark sector, but there the dominant suppression is coming from QCD enhancement of the diagonal squark masses.
      
  Turning to the less statistically significant evidence for $B_d$ mixing, comparing the SUSY predictions in 
  Eq.\ref{SUSY} to the best fit points in Eq.\ref{bestfit}, it is clear that the values of $h_d \sim 0.25$ 
  could either be achieved by $(\delta^d_{31})_{LL}\sim (\delta^d_{31})_{RR} \sim 0.8\times 10^{-2}$
or $(\delta^d_{31})_{LL}\ll (\delta^d_{31})_{RR} \sim 5\times 10^{-2}$ or 
 $(\delta^d_{31})_{RR}\ll (\delta^d_{31})_{LL} \sim 5\times 10^{-2}$.  From the point of view of
 conventional family $SU(3)$ symmetry models the above squark mixings are much larger than the expected values  $(\delta^d_{31})_{RR}\sim 10^{-4}$  \cite{Antusch:2008jf}.  Although these low energy squark mixings
 look more modest, they must originate from high scale squark mixings which are many times larger than these values (25 times if the high scale is the GUT scale).
 Again we consider the smallest mixing case corresponding to 
 $(\delta^d_{31})_{LL}\sim (\delta^d_{31})_{RR} \sim 0.8\times 10^{-2}$ corresponding to high scale values perhaps of order $\lambda \sim 0.2$. This suggests a model with (1,3) quark mixing angles of order $\lambda$ in the up and down sectors, 
 $\theta^u_{13} \sim \theta^d_{13} \sim \lambda$ as compared to 
 $\theta_{13}\sim \lambda^3$, which again demands some natural cancellation mechanism. 
  In order to achieve this we may extend the $SU(3)$ approach above, by introducing
 using the anti-triplet flavon $\phi_{123}^i$ with aligned VEV
 $\langle \phi_{123} \rangle \sim (1, 1, 1)V_{123}$, where phases    
 have been suppressed (see \cite{Antusch:2008jf}), which introduces the additional operators
\be
 H^uQ_i\phi_{123}^i   U^c_j\phi_{23}^j   +  H^uQ_i\phi_{23}^i   U^c_j\phi_{123}^j  
  + H^dQ_i\phi_{123}^i   D^c_j\phi_{23}^j  + H^dQ_i\phi_{23}^i   D^c_j\phi_{123}^j .
  \label{o4}
   \ee 
 The combined effect of the operators in Eqs.\ref{o1},\ref{o4} is to yield the quark Yukawa couplings,
 assuming $V_{123}/V_{23} \sim \lambda$, ignoring phases,
 \be
Y^u \sim Y^d \sim \left(
  \begin{matrix} % or pmatrix or bmatrix or Bmatrix or ...
  0 & \lambda & \lambda \\
\lambda &  1 + 2\lambda & 1 + 2 \lambda\\
\lambda  &  1 + 2 \lambda & 1 + 2\lambda \\
   \end{matrix}
   \right).
\ee  
After the maximal angle (2,3) rotations, the remaining matrices are diagonalised by separate and equal (1,3) rotations
of order $\lambda$ in both the up and down sectors which cancel as required. 
However this results in first family eigenvalues of order 
$\lambda^2$ in both the up and down sectors leading to the relations $m_u/m_t = m_d/m_b \sim \lambda^2$
which are in strong disagreement with their observed values. This result could in principle be evaded when 
other operators or corrections are included as required to describe
the second family quark masses $m_c,m_s$, but apparently only at the expense of 
an unnatural cancellation of the leading order first family result above.
Therefore we are not able
to accommodate large and natural (1,3) mixing of order $\lambda$ in the up and down sectors as suggested by the 
$B_d$ mixing fits. 
In general we do not see how to achieve this in a natural way in the MSSM including 
family symmetry and GUTs.  We conclude that either the weaker requirement of non-standard $B_d$ mixing be discarded, which as remarked in the Introduction is only required at a confidence level of order 1.5$\sigma$, or one must abandon the MSSM as a natural explanation for CP violation in both 
$B_s$ and $B_d$ mixing within the framework of family symmetries.

  \section{Large CP Violation for $B_s$ and $B_d$ mixing with three families of SUSY Higgs} 
It is obviously easier to account for large CP violation in $B_s$ and $B_d$ mixing by the tree-level exchange of some new heavy particles than if they only appear in a one-loop diagram, as is the case for squarks and gluinos in the MSSM which only appear in the SUSY box diagram considered in the previous section. 
Indeed many of the approaches suggested in  \cite{Dighe:2010nj,Dobrescu:2010rh,Chen:2010wv,Buras:2010mh} are based the tree-level exchanges of some new particles. 

A good example of this approach is the suggestion \cite{Dobrescu:2010rh} that CP violation in $B_s$ mixing may be due to the tree-level exchange of a new neutral spin-0 boson 
$H_{\mathrm{d}}^0=(H^0+iA^0)/\sqrt{2}$ which is the neutral component of a Higgs doublet where $H^0$ does not get a VEV at leading order so that $H^0$ and $A^0$ have approximately the same mass $M_H=M_A$. The Yukawa couplings of $H_{\mathrm{d}}^0$ to $b,s$ quarks in the mass eigenstate basis are given  \cite{Dobrescu:2010rh} ,
\be
-H_{\mathrm{d}}^0(y_{bs}\overline{b}_Rs_L + y_{sb}\overline{s}_Rb_L   ) + H.c.
\label{ybs}
\ee
Tree-level $H_{\mathrm{d}}^0$ exchange gives rise to the following operator which contributes to $B_s$ mixing,
\be
\frac{y_{bs}y_{sb}^*}{M_H^2}(\overline{b}_Rs_L )(\overline{b}_Ls_R).
\ee

%These couplings then give rise to $B_s$ mixing and CP violation at the Tevatron as depicted in 
%Fig.\ref{fig0}.
%\begin{figure}[htbp]
%\begin{center}
%\includegraphics[scale=0.4]{fig0.pdf}
%\caption{Mechanism for $B_s$ mixing and CP violation at the Tevatron due to new Higgs exchange
%\cite{blog}.}
%\label{fig0}
%\end{center}
%\end{figure}

One may readily extend this idea to allow for couplings of $H_{\mathrm{d}}^0$ to $b,d$ quarks in the mass eigenstate basis,
\be
-H_{\mathrm{d}}^0(y_{bd}\overline{b}_Rd_L + y_{db}\overline{d}_Rb_L   ) + H.c.
\label{ybd}
\ee
Tree-level $H_{\mathrm{d}}^0$ exchange then gives rise to the following operator which contributes to $B_d$ mixing,
\be
\frac{y_{bd}y_{db}^*}{M_H^2}(\overline{b}_Rd_L )(\overline{b}_Ld_R).
\ee

Following the approach of the previous section, the $B_q$ mixing matrix element can be written as the sum of the SM box diagrams and the Higgs tree-level exchange diagrams,
\be
M^{12}_q = M^{12,\mathrm{SM}}_{q} + M^{12,\mathrm{Higgs}}_{q}
\ee
where from Eq.\ref{h} we identify
\be
h_qe^{i \theta_q}=\frac{ M^{12,\mathrm{Higgs}}_{q}}{M^{12,\mathrm{SM}}_{q}},
\ee
where $q=s,d$.
Using the matrix elements in \cite{Buras:2010mh} we find,
\bea
h_se^{i \theta_s} & \approx &  -1.4\times 10^6y_{bs}y_{sb}^*
 \left( \frac{200\ \mathrm{GeV}}{M_H}\right)^2
\nonumber \\
h_de^{i (\theta_d + 2\beta)} & \approx &  -3.4\times 10^7y_{bd}y_{db}^*
 \left( \frac{200\ \mathrm{GeV}}{M_H}\right)^2.
\label{Higgs}
\eea
Comparing the Higgs predictions in Eq.\ref{Higgs} to the best fit points in Eq.\ref{bestfit}, it is clear that the values of $h_s \sim 1$ could be achieved, assuming Higgs masses of 
about 200 GeV,  by $y_{bs}\sim y_{sb}\sim 10^{-3}$, depending on the phases.
Similarly, $h_d \sim 0.25$ could be achieved, assuming Higgs masses of 
about 200 GeV,  by $y_{bd}\sim y_{db}\sim 10^{-4}$, depending on the phases.
In both cases the required Yukawa couplings are proportional to the Higgs mass.

Of course such tree-level exchanges must be kept under control so that they don't induce too much flavour changing in other places where the constraints are more severe, especially involving the first two quark and lepton families as is the case for example with $\epsilon_K$.
To overcome these challenges, it has been suggested that the hypothesis of Minimal Flavour Violation (MFV) could be extended to the two (or more) Higgs doublet model \cite{Buras:2010mh}. Here we shall follow a different approach, namely to use the idea of family symmetry to control the magnitude of the flavour changing Yukawa couplings in the framework of a model with three families of SUSY Higgs doublets. 

The basic idea we shall discuss is very simple, namely that there are three SUSY Higgs families which form triplets under some family symmetry group, just like the three families of quarks and leptons. The three SUSY Higgs families can be written as $H_i^u,H_i^d$ where the index $i=1,2,3$ labels the three copies of the two MSSM Higgs doublets $H^u,H^d$. The idea is that the three families of quarks $Q_i,U^c_i,D^c_i$  and Higgs $H_i^u,H_i^d$ all transform as triplets $\sim 3$ under some family symmetry group. Yukawa couplings are forbidden in the exact family symmetry limit but are allowed when the family symmetry is broken by flavon VEVs. The idea is that the family symmetry leads to a doubly hierarchical structure as follows.
The third family of Higgs couples more strongly to quarks and leptons than the first and second Higgs families. The third family of Higgs also couples more strongly to the third family of quarks than the first and second families of quarks. This leads to the observed pattern of quark masses and mixings due to the third Higgs family couplings, with suppressed flavour changing couplings from the first and second Higgs families.

A model with three families of SUSY Higgs which controls the flavour changing using a
family symmetry has recently been discussed \cite{Howl:2009ds}.
The model \cite{Howl:2009ds} is based on a $\Delta_{27}$
family symmetry and leads to a successful 
description of all quark and lepton masses and mixing, including tri-bimaximal lepton mixing
\cite{deMedeirosVarzielas:2006fc}.
However only the dominant third Higgs family couplings were considered \cite{Howl:2009ds}
and flavour and CP violation arising from the subdominant first and second family Higgs couplings were not considered. In the following we shall revisit this model, focussing on the quark sector,
assuming exactly the same particle content and symmetries as in \cite{Howl:2009ds}, but
including the effects of different field orderings and contractions not previously considered.

In the considered model \cite{Howl:2009ds} 
the quarks and Higgs transform as triplets under a  $\Delta_{27}$ family symmetry, which is broken by anti-triplet flavons $\phi_{3},\phi^h_{3},\phi_{23},\phi_{123} $
which develop aligned VEVs, dropping phases in the following discussion for simplicity
(they are recovered in Appendix~\ref{A}),
\begin{equation} \label{eq:VEVs}
\langle \phi_{3}  \rangle , \langle \phi^h_{3}  \rangle \propto 
\left( \begin{array}{ccc} 0 & 0 & 1 \end{array} \right)^T,\
\langle \phi_{23} \rangle \propto \left( \begin{array}{ccc} 0 & 1 & 1 \end{array} \right)^T,\
\langle  \phi_{123}\rangle  \propto \left( \begin{array}{ccc} 1 & 1 & 1 \end{array} \right)^T.
\end{equation}
The leading order down-type quark Yukawa couplings result from the following flavon couplings, suppressing coupling constants,
\bea
\frac{1}{M_d^2M_h} &[&  (Q \phi_3)_1 (D^c \phi_3)_1 (H^d \phi^{h}_3)_1 +
(Q \phi_{23})_1 (D^c \phi_{23})_1 (H^d \phi^{h}_3)_1  \nonumber \\
&+& (Q \phi_{123})_1 (D^c \phi_{23})_1 (H^d \phi^{h}_3)_1  + 
(Q \phi_{23})_1 (D^c \phi_{123})_1 (H^d \phi^{h}_3)_1 \ \  ] , \label{Howl}
\eea
with similar couplings generating the up-type and lepton Yukawa couplings.
The vacuum alignments in Eq.\ref{eq:VEVs} then imply that only the third Higgs family couples to quarks and leptons with the leading order Yukawa matrix, 
defined by $Y^{ij3}_dQ_iD^c_jH^d_3$, given as,
\be Y^{ij3}_d \sim \left( \begin{array}{ccc} 0 &  \epsilon^3_d &  \epsilon^3_d
\\ \epsilon^3_d & \epsilon^2_d & \epsilon^2_d
\\  \epsilon^3_d &  \epsilon^2_d & 1 \end{array} \right)  
\label{Yd3}
\ee
where the expansion parameter $\epsilon_d \approx 0.15$, where 
we have assumed \cite{Howl:2009ds} 
\footnote{More precisely 
$\frac{\langle\phi^h_{3}\rangle}{M_h} \approx \frac{\langle\phi_{3}\rangle}{M_d} \approx 0.8$
\cite{Howl:2009ds} .},
\be
\frac{\langle\phi^h_{3}\rangle}{M_h} \approx
\frac{\langle\phi_{3}\rangle}{M_d} \sim 1, \ \
\frac{\langle\phi_{23}\rangle}{M_d} \approx \epsilon_d, \ \
\frac{\langle\phi_{123}\rangle}{M_d}
\approx \epsilon^2_d.
\label{exp1}
\ee 
A similar Yukawa matrix arises for the up-type quarks but with a smaller expansion
parameter $\epsilon_u \approx 0.05$.
When supplemented by additional corrections \cite{Howl:2009ds}, 
such Yukawa matrices, after the third Higgs family develop their VEVs, have been shown to provide a successful description of quark masses and mixing \cite{Ross:2007az}.

\begin{figure}[htbp]
\begin{center}
\includegraphics[scale=0.4]{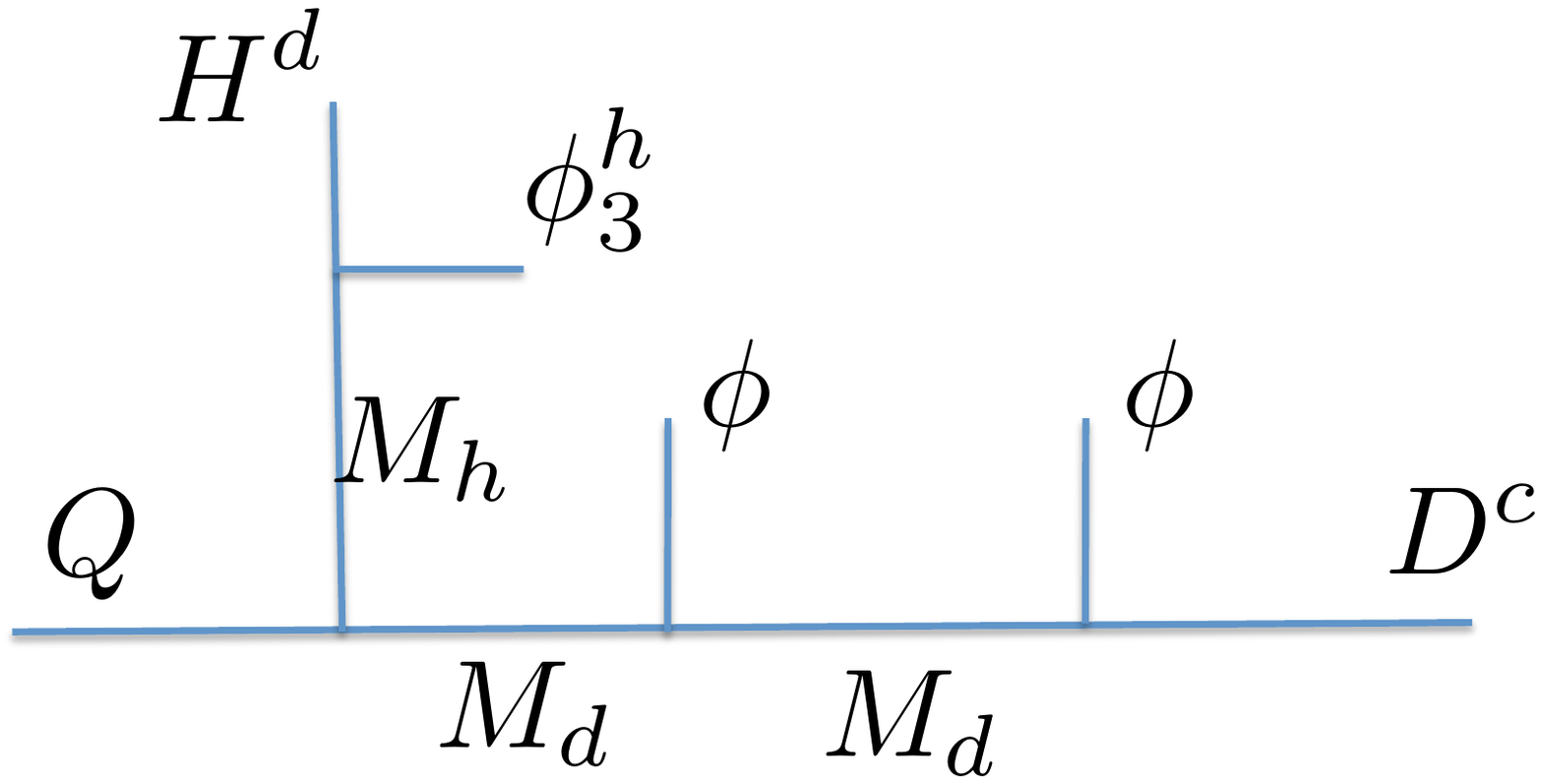}
\caption{The messenger diagram for the down-type quark sector as assumed in \cite{Howl:2009ds} where $\phi$ represents any of the flavons $\phi_3, \phi_{23},\phi_{123}$ allowed by the symmetry. Analogous diagrams are present in the up-type quark and lepton sectors.}
\label{fig1}
\end{center}
\end{figure}

The couplings in Eq.\ref{Howl} are controlled by the additional symmetries of \cite{Howl:2009ds} and
are generated by the exchange of heavy messenger particles of mass $M_{d}$ and $M_h$,
as indicated in Fig.\ref{fig1}. The quark singlet messengers with masses $M_{d}$ 
are assumed to be lighter than the quark doublet messengers of mass $M_Q$, and hence give the dominant contribution with $M_{u}\approx 3M_{d}$
being responsible for $\epsilon_d \approx 3 \epsilon_u$.
The assumed messenger sector also ensures that $\phi^{h}_3$ couples 
directly to the Higgs doublets with the group theory contractions giving the $\Delta_{27}$
singlet as indicated by the subscript in $(H^d \phi^{h}_3)_1$. This 
implies that only the third Higgs family couples to quarks.

\begin{figure}[htbp]
%\begin{center}
\includegraphics[scale=0.3]{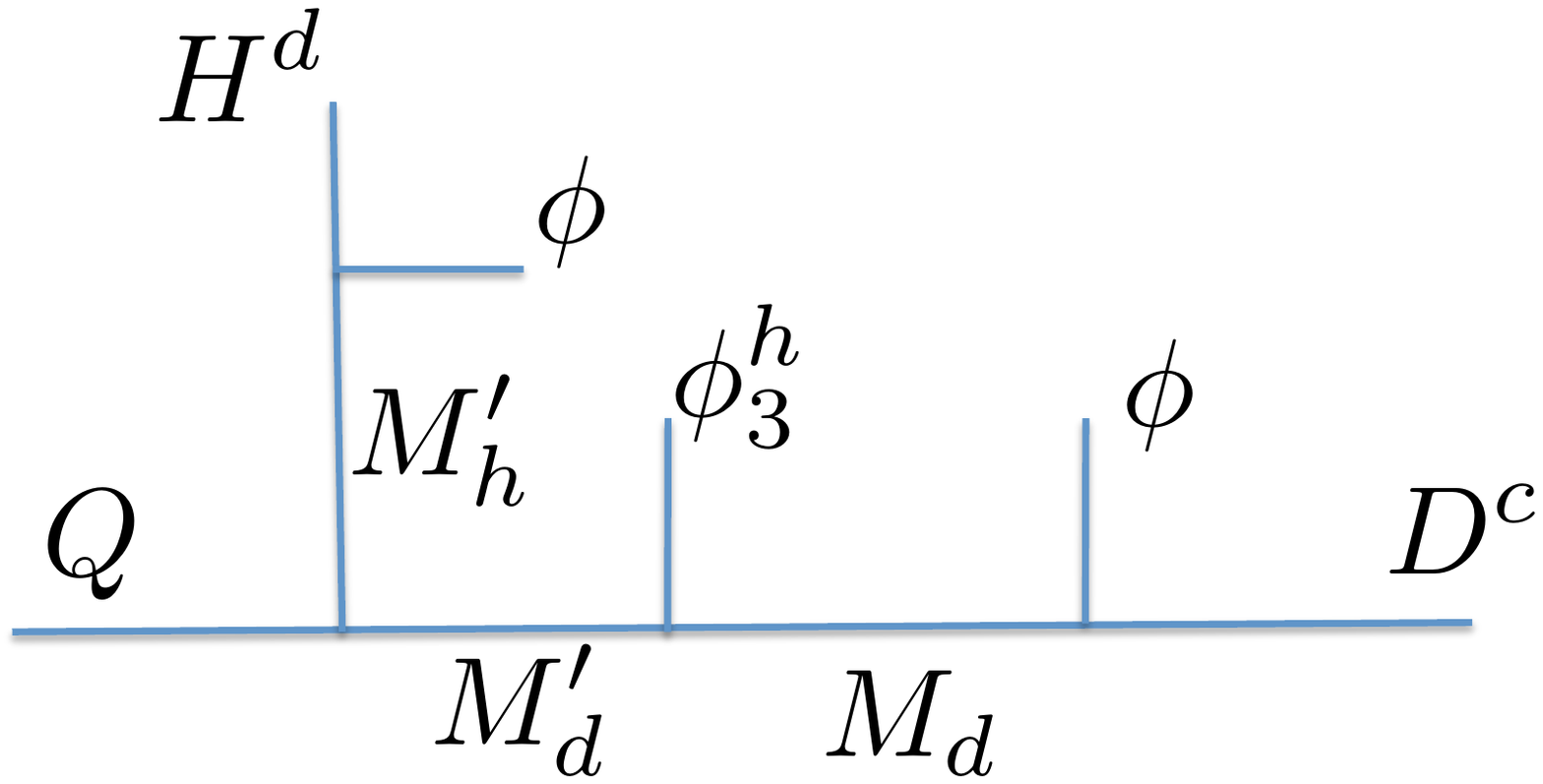}
\includegraphics[scale=0.3]{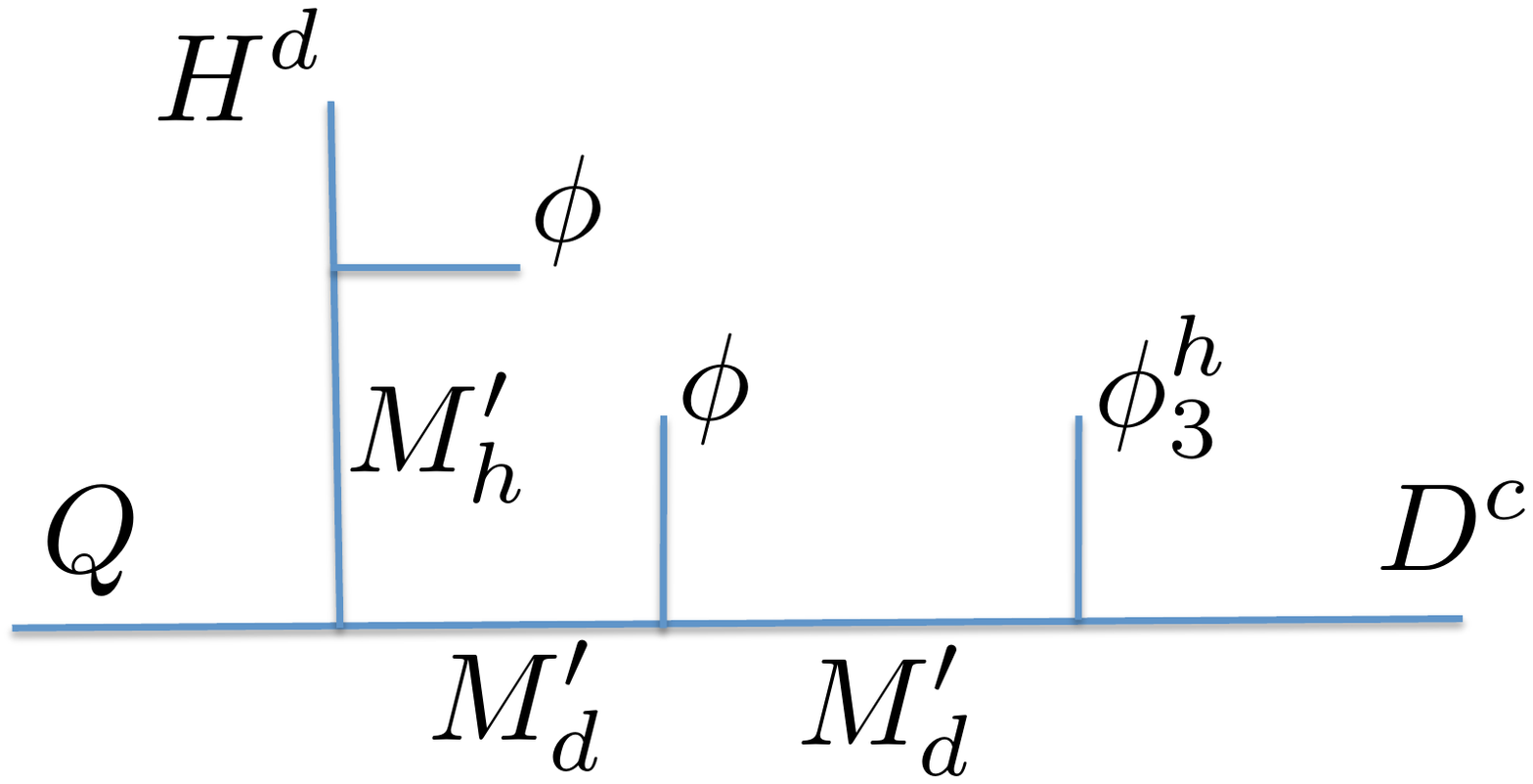}
\caption{Other possible messenger diagrams involving the same flavons as in 
Fig.\ref{fig1}. Additional diagrams with two flavons along the Higgs line are also possible but we assume that
diagrams with three flavons along the same line are suppressed due to higher dimensional messenger
representations being  absent or very heavy.
The primed messenger masses are assumed to larger than the unprimed ones,
leading to the diagram on the left being suppressed compared to Fig.\ref{fig1},
and the diagram on the right being even more suppressed.}
\label{fig2}
%\end{center}
\end{figure}

In  Fig.\ref{fig2} we display other
messenger diagrams, not considered in \cite{Howl:2009ds}, allowed by all the symmetries of the model and
involving the same flavon fields but contracted differently. These operators will
contribute at a suppressed level due to the (assumed) heavier primed messenger masses. 
The diagram on the left in Fig.\ref{fig2} will lead to the following additional operators, suppressed by two primed messenger masses,
\bea
\frac{1}{M_dM_d'M_h'} &[&  (Q \phi^{h}_3)_1 (D^c \phi_3)_1 (H^d\phi_3)_1 +
(Q  \phi^{h}_3)_1 (D^c \phi_{23})_1 (H^d\phi_{23})_1  \nonumber \\
&+& (Q \phi^{h}_3)_1 (D^c \phi_{23})_1 (H^d\phi_{123})_1 +
(Q \phi^{h}_3 )_1 (D^c \phi_{123})_1 (H^d\phi_{23})_1 \ \  ] , \label{left}
\eea
with similar couplings for up-type quarks and leptons.
The diagram on the right in Fig.\ref{fig2} will lead to the following additional operators, suppressed by 
three primed messenger masses,
\bea
\frac{1}{{M_d'}^2M_h'} &[&  (Q \phi_3)_1 (D^c\phi^{h}_3 )_1 (H^d \phi_3)_1 +
(Q \phi_{23})_1 (D^c \phi^{h}_3)_1 (H^d\phi_{23})_1  \nonumber \\
&+& (Q \phi_{123})_1 (D^c  \phi^{h}_3)_1 (H^d\phi_{23})_1  +
(Q \phi_{23})_1 (D^c  \phi^{h}_3)_1 (H^d\phi_{123})_1 \ \  ] , \label{right}
\eea
with similar couplings for up-type quarks and leptons.
Other messenger diagrams with two flavons along the Higgs line will not introduce new
operator structures but will only change the overall coefficient
of the operators in Eq.\ref{left},\ref{right} with one power of $M_d'$ being replaced by $M_h'$. 
Additional operator structures are also present 
in which the singlet contractions (represented by the subscripts)
are replaced by the one dimensional representations $\mathbf{1'},\mathbf{1''}$
present in  $\Delta_{27}$, as discussed in Appendix~\ref{A}. However we assume that messengers
in other one dimensional or higher dimensional representations are absent or very heavy.

The new couplings in Eqs.\ref{left},\ref{right}, together with 
the vacuum alignments in Eq.\ref{eq:VEVs}, allow the first and second 
Higgs families to couple to quarks and leptons with the leading order Yukawa matrices,
defined by $Y^{ijk}_dQ_iD^c_jH^d_k$, given as,
\be 
Y^{ij2}_d \sim \left( \begin{array}{ccc} 0 &  0  &  \epsilon^3_d\alpha_d
\\ 0  & 0  & \epsilon^2_d \alpha_d \\  
\epsilon^3_d &  \epsilon^2_d &  \epsilon^2_d \end{array} \right)\alpha_d \alpha_h, \ \ \ \ 
Y^{ij1}_d \sim \left( \begin{array}{ccc} 0 &  0  &  0 \\ 
0 & 0 & \epsilon^3_d \alpha_d
\\  0 &  \epsilon^3_d &  \epsilon^3_d \end{array} \right)\alpha_d \alpha_h,
\label{Yd1}
  \ee
where we have used the expansion parameters in Eq.\ref{exp1}, together with the 
suppression factors,
\be
\alpha_d\approx \frac{M_d}{M_d'}, \ \ \alpha_h\approx \frac{M_h}{M_h'}.
\label{exp2}
\ee
Eq.\ref{Yd1} shows that the second family Higgs $H^d_2$ 
couples more strongly to quarks than the
first family Higgs $H^d_1$ whose effects can be approximately ignored.   
Assuming as a leading order approximation 
that only the third Higgs family develops VEVs, which is reasonable since the VEVs are
radiatively generated as a result of the large third family Yukawa coupling
in Eq.\ref{Yd3} of order unity,  
then we may diagonalise the quark mass matrix
resulting from Eq.\ref{Yd3} by small angle rotations $\theta^d_{23}\sim \epsilon_d^2$,
$\theta^d_{13}\sim \epsilon_d^3$, $\theta^d_{12}\sim \epsilon_d$,
to go to the diagonal down-type quark mass basis $d,s,b$. 
The neutral component of the second family Higgs ${H^d_2}^0$ then has a Yukawa coupling
matrix (the first equation in Eq.\ref{Yd1}) which, when rotated to the down quark mass basis,
takes the leading order form, 
\be 
Y^{ij2}_{d_\mathrm{mass}} \sim \left( \begin{array}{ccc} \epsilon^6_d &  \epsilon^5_d  &  \epsilon^3_d\alpha_d
\\   \epsilon^5_d &  \epsilon^4_d  & \epsilon^2_d \alpha_d \\  
\epsilon^3_d &  \epsilon^2_d &  \epsilon^2_d \end{array} \right)\alpha_d \alpha_h,
\label{Yd2}
  \ee
in the convention where the rows correspond to
$d_L,s_L,b_L$ and the columns correspond to $d_R,s_R,b_R$.  
Clearly the flavour violating couplings of ${H^d_2}^0$ involving $b_R$ in Eq.\ref{Yd2}
are relatively suppressed by a factor of $\alpha_d$
compared to those involving $b_L$, as can be understood from Fig.\ref{fig2}.
Note that there are no cancellations of the couplings in Eq.\ref{Yd2}
 in the $d,s,b$ basis since each operator
in Eqs.\ref{Howl},\ref{left},\ref{right} has an independent order unity coefficient which has been suppressed
for clarity. The flavour violating couplings in Eqs.\ref{ybs},\ref{ybd} can then be read off from 
Eq.\ref{Yd2}, where we identify $H_{\mathrm{d}}^0\equiv {H^d_2}^0$,
\be
y_{bs}\sim  \epsilon^2_d \alpha_d^2 \alpha_h, \ \ 
y_{sb}\sim  \epsilon^2_d \alpha_d \alpha_h, \ \ 
y_{bd}\sim  \epsilon^3_d \alpha_d^2 \alpha_h, \ \ 
y_{db}\sim  \epsilon^3_d \alpha_d \alpha_h.
\label{y}
\ee

With the phases included, as discussed in Appendix~\ref{A},
the couplings in Eq.\ref{y} lead to CP violation with,
\be
\frac{|y_{bd}y_{db}^*|}{|y_{bs}y_{sb}^*|}\sim \epsilon_d^2 \sim 2\times 10^{-2},
\ee
which leads to the prediction, using Eq.\ref{Higgs},
\be
\frac{h_d}{h_s}\sim \frac{1}{2},
\ee
in good agreement with the best fit point I in Eq.\ref{bestfit}.
This prediction is independent of the messenger masses and hence is equally valid
for other messenger diagrams with two flavons along the Higgs line.

Assuming best fit point I the required Higgs mass may be related to 
$\alpha_d,\alpha_h$ using Eq.\ref{Higgs} , 
\be
 \left( \frac{M_H}{200\ \mathrm{GeV}}\right)^2
 \approx 10^3\alpha_d^3 \alpha_h^2.
 \ee
For example $\alpha_d \sim \alpha_h \sim 1/3$ would require $M_H \sim 400$ GeV.
This choice of couplings would then imply,
\be 
y_{sb}\sim  \epsilon^2_d \alpha_d \alpha_h \sim 2.5\times 10^{-3}, \ \ 
y_{db}\sim  \epsilon^3_d \alpha_d \alpha_h \sim 3.7\times 10^{-4}.
\label{ys}
\ee

The hierarchical structure of flavour changing couplings in Eq.\ref{Yd2} suppresses the contribution
of second family Higgs exchange to $K^0$ mixing, which is described by a very small Yukawa coupling
\be
y_{sd}\sim  y_{ds}\sim  \epsilon^5_d \alpha_d \alpha_h \sim 8\times 10^{-6},
\label{yd}
\ee
where we have assumed $\alpha_d \sim \alpha_h \sim 1/3$. 
The charged leptons are expected to have flavour violating 
couplings with a similar structure to  Eq.\ref{Yd2}, with the hierarchical structure again leading to 
suppressed lepton flavour violating processes with
\be
 y_{\mu e} \sim y_{e \mu} \sim y_{sd}\sim  y_{ds} \sim 8\times 10^{-6}.
  \label{ye}
\ee
The diagonal coupling of the second family Higgs to muons is also quite suppressed,
 \be
 y_{\mu \mu} \sim  \epsilon^4_d \alpha_d \alpha_h \sim 6\times 10^{-5},
\label{ymu}
\ee
leading to a negligible contribution to $Br (B_s \rightarrow \mu^+ \mu^-)$. 
For example, assuming $\epsilon_d \approx 0.15$, 
$\alpha_d \sim \alpha_h \sim 1/3$ and $M_H \sim 400$ GeV using the results in 
\cite{Dobrescu:2010rh} we estimate that the contribution 
from second family Higgs exchange to the branching ratio is 
$\Delta Br (B_s \rightarrow \mu^+ \mu^-) \approx 3 \times 10^{-13}$ which is negligible
compared to the SM prediction of about $3.7 \times 10^{-9}$.
This contrasts with 
other non-standard Higgs models \cite{Dobrescu:2010rh,Buras:2010mh} which tend to predict
$Br (B_s \rightarrow \mu^+ \mu^-)$ larger than the SM value and close to the current experimental limit
of about $5.8 \times 10^{-8}$.

Similarly the diagonal coupling of the second family Higgs to taus is,
 \be
 y_{\tau \tau} \sim  \epsilon^2_d \alpha_d \alpha_h \sim 2.5\times 10^{-3},
\label{ytau}
\ee
leading to a new contribution, assuming the same parameters as above, 
$\Delta Br (B_s \rightarrow \tau^+ \tau^-)  \approx 4.5 \times 10^{-10}$, which is also somewhat 
below the SM prediction of about $2.7 \times 10^{-9}$, including a phase
space suppression factor of 0.75 in both cases. However, a different choice of parameters
could enhance the new physics contribution and make it competetive with the SM contribution.

It is worth recalling that in the SM the decay matrix element $\Gamma_{12}^s$ is proportional to the square of a tree-level amplitude proportional to $V_{cb}\sim 4\times 10^{-2}$ arising from $W$ exchange. 
As remarked in the Introduction, the best fit value for $\Gamma_{12}^s$ is about twice as large
as the SM prediction and it is challenging to understand this. For example, in the present model,
second family Higgs exchange with mass $M_H \sim 400$ GeV and couplings in 
Eqs.\ref{ys},\ref{yd},\ref{ye},\ref{ymu},\ref{ytau} give a contribution to $\Gamma_{12}^s$
which is completely negligible compared to the SM 
$W$ exchange contribution. The corresponding charged Higgs exchange contributions are also expected to be suppressed compared to the SM. For example, the interaction 
$y_{cb}H_{\mathrm{d}}^+\overline{c}_Rb_L $ involves a coupling,
\be 
y_{cb}\sim y_{sb}\sim  \epsilon^2_d \alpha_d \alpha_h \sim 2.5\times 10^{-3},
\label{ycb}
\ee
which is again smaller than $V_{cb}\sim 4\times 10^{-2}$, with all charged Higgs couplings involving at least this suppression, and the charged Higgs mass being heavier than the $W$ mass.

Finally we remark that the model in \cite{Howl:2009ds}, as developed above, is based on  
$\Delta_{27}$ family symmetry combined with the
E$_6$SSM  \cite{King:2005jy} which predicts three complete SUSY Higgs families as part of three complete 27 dimensional SUSY matter representations at the TeV scale (minus three right-handed neutrinos which get high see-saw scale masses since they carry no charges in this model). In addition there is a pair of SUSY doublets 
$L, \overline{L}$ in conjugate representations
which form a TeV scale Dirac mass as required for GUT scale unification. These may be absent if the requirement of GUT scale unification is relaxed \cite{Howl:2007zi}. The E$_6$SSM can also be tested via its prediction of a $Z'_N$ gauge boson with flavour conserving couplings \cite{King:2005jy,Howl:2007zi}.

\section{Conclusion}
Following the anomalous like-sign dimuon charge asymmetry measured by the D0 collaboration 
at the Tevatron collider we have discussed the implications of large CP violation in $B_{d,s}$ mixing for Supersymmetric (SUSY) Standard Models, focussing on those models which allow a family symmetry and unification. For the Minimal Supersymmetric Standard Model (MSSM) we have seen
from Eq.\ref{SUSY} that it is only possible to account for $B_{s}$ mixing and CP violation at the expense of large squark mixing which would require a 
new approach to family symmetry models. However, assuming such a framework, it seems very difficult to account
for a significant amount of $B_{d}$ mixing and CP violation.
 
In order to describe both $B_{s}$ and $B_{d}$ mixing and CP violation, as suggested by a recent global fit, we were led to consider SUSY models with Higgs fields transforming as triplets under a family symmetry. We have described a realistic such model based on $\Delta_{27}$ family symmetry combined with the 
E$_6$SSM in which tree-level exchange of the second Higgs family predicts $B_{s}$ and $B_{d}$ mixing and CP violation in the ratio $h_d/h_s \sim 1/2$ which is in good agreement with best fit point I in Eq.\ref{bestfit}. The model naturally suppresses flavour and CP violation in $\epsilon_K$
and the lepton sector, and is distinguished from other Higgs models by predicting
$Br (B_s \rightarrow \mu^+ \mu^-)$ consistent with the SM prediction.  
  
   \section*{Acknowledgements}
We would like to thank J.~Flynn, D.~King, C.~Luhn and R.~Zwicky for useful discussions.
We acknowledge the support of a Royal Society Leverhulme Trust Senior Research Fellowship and the STFC Rolling Grant ST/G000557/1.
 
 \section*{Appendix}
\appendix
\section{The origin of phases}\label{A}

In this Appendix we discuss the question of the origin of the
phases in the flavour violating Yukawa couplings
in Eq.\ref{y}. Following the approach in \cite{Antusch:2008jf}, we shall assume that CP is preserved in the high energy theory but is 
spontaneously broken by the flavon VEVs in Eq.\ref{eq:VEVs} whose 
phases can be restored as follows,
\bea \label{phases}
\langle \phi_{3}  \rangle & \propto &  
\left( \begin{array}{ccc} 0 & 0 & e^{i\omega_3} \end{array} \right)^T, \ \ 
\langle \phi^h_{3}  \rangle \propto 
\left( \begin{array}{ccc} 0 & 0 & e^{i(\omega_3+\phi_h)} \end{array} \right)^T,\nonumber \\
\langle \phi_{23} \rangle & \propto &  
\left( \begin{array}{ccc} 0 &  e^{i\omega_2} &   e^{i(\omega_3+\phi_3)}
\end{array} \right)^T,\
\langle  \phi_{123}\rangle  \propto 
\left( \begin{array}{ccc}  e^{i\omega_1} &  e^{i(\omega_2+\phi_1)} &  e^{i(\omega_3+\phi_2)}
\end{array} \right)^T,
\eea
where the phases $\omega_i$ can be removed by $SU(3)$ transformations but not
in the $\Delta_{27}$ theory. Another difference between $SU(3)$ and 
$\Delta_{27}$ is that the discrete symmetry allows nine distinct one dimensional
representations
\cite{Ma:2006ip}, which, depending on the messenger representations, allows many more new operators than those given in Eqs.\ref{left}, \ref{right}, corresponding to the different singlet contractions $(\mathbf{3}\times \overline{\mathbf{3}} )_{\mathbf{1_r}}$ where $r=1,\ldots 9$. Here we restrict ourselves to 
$A_4$ type messengers in the first three one dimensional representations 
which can be obtained from the products $\mathbf{3}\times \overline{\mathbf{3}}$ as follows,
\be
\mathbf{1}=1\overline{1} + 2\overline{2} + 3\overline{3}, \ \ 
 \mathbf{1'}=1\overline{1} + \omega 2\overline{2} + \omega^2 3\overline{3}, \ \ 
 \mathbf{1''}=1\overline{1} + \omega^2 2\overline{2} + \omega 3\overline{3}, \ \ 
   \ee
which are familiar from $A_4$ \cite{Ma:2001dn} where $\omega = \mathrm{exp}(2\pi i /3)$.
Allowing messengers in the $\mathbf{1},\mathbf{1'},\mathbf{1''}$ representations 
permits new operators corresponding 
to one messenger in each of the allowed one dimensional representations, where the invariant singlet is given by $\mathbf{1}= \mathbf{1'} \times \mathbf{1''}$. Thus the operators in Eqs.\ref{left}, \ref{right} need to be augmented by others of the more general form,
$(\cdots )_{\mathbf{1}}(\cdots )_{\mathbf{1'}}(\cdots )_{\mathbf{1''}}$ appearing in all possible
combinations. 
Assuming the messengers in the 
$\mathbf{1},\mathbf{1'},\mathbf{1''}$ representations all have the same mass,
the expansion parameters and predictions given previously will not change.
In particular the flavour violating couplings will have their magnitudes unchanged
from the values quoted in Eq.\ref{y}, but their phases will all be different from each other
in a complicated way which depends on the order unity couplings which control
the precise linear combinations of the different operators which contribute to these
couplings. In the limit that only singlet operators are permitted 
$(\cdots )_{\mathbf{1}}(\cdots )_{\mathbf{1}}(\cdots )_{\mathbf{1}}$ 
it is easy to show that $\mathrm{arg}(y_{bs}) = \mathrm{arg}(y_{sb})$ and 
 $\mathrm{arg}(y_{bd}) = \mathrm{arg}(y_{db})$, even with the most general
 flavon VEVs in Eq.\ref{phases}, so the extra operators of the form 
 $(\cdots )_{\mathbf{1}}(\cdots )_{\mathbf{1'}}(\cdots )_{\mathbf{1''}}$
are in fact necessary in order to allow new sources of CP violation in $B_s$ and $B_d$ mixing.

\end{document}